\documentclass[aip,jap,groupedaddress,reprint]{revtex4-1} 
\usepackage{amsmath,amssymb,graphicx,MnSymbol}
\usepackage{booktabs}
\usepackage{multirow}
\usepackage{rotating}
\newcommand{\sss}{\scriptscriptstyle}
\newcommand{\sst}{\scriptstyle}

\newcommand{\stext}[1]{\sss \text{#1} \sst}

\begin{document}

\title{Anisotropic Bruggeman Effective Medium Approaches for Slanted Columnar Thin Films}

\author{Daniel Schmidt}\email{schmidt@engr.unl.edu}
\author{Mathias Schubert}
\affiliation{University of Nebraska-Lincoln, Department of Electrical Engineering, Lincoln, NE 68588, USA}

\begin{abstract}
Two different formalisms for the homogenization of composite materials containing ellipsoidal inclusions based on Bruggeman's original formula for spherical inclusions can be found in the literature. Both approximations determine the effective macroscopic permittivity of such an idealized composite assuming randomly distributed dielectric particles of equal shape and differ only in the definition of the depolarization factors. The two approaches are applied to analyze ellipsometric Mueller matrix spectra acquired in the visible and near-infrared spectral region from metal and semiconductor slanted columnar thin films. Furthermore, the effective dielectric function tensor generated by the two Bruggeman formalisms is compared to effective major axes dielectric functions individually determined with a homogeneous biaxial layer approach. Best-match model parameters of all three model approaches are discussed and compared to estimates from scanning electron microscope images.
The structural parameters obtained from all three optical modeling approaches agree well with the electron microscopy technique. A comparative discussion is given for the validity and applicability of the three model approaches for analysis of future devices structures that may require optical readout using generalized ellipsometry methods.
\end{abstract}


\maketitle 


\section{Introduction}
Functionalizing thin films by controlled porosity engineering is a very interesting route to achieve desired physical film properties for various applications such as chromatography or optical humidity and biochemical sensing~\cite{Jim2011,Liu2011,Rodenhausen2012}. Glancing angle deposition is an appropriate physical vapor deposition technique to fabricate nanostructured thin films with columnar characteristics and allows for tailoring especially optical properties by controlling nanostructure geometry and porosity~\cite{Hawkeye2007}. Evaluating structural and optical properties of these three-dimensional columnar thin films to optimize deposition conditions and to improve device properties is desirable. Scanning probe microscopy and scanning electron microscopy (SEM) techniques are suitable for determining surface morphological and structural properties such as column and film thickness and columnar titling angles~\cite{Backholm2012}. Obtaining reliable estimates about film porosity (void fraction) can be rather difficult and typically top-down SEM images are analyzed; however, this technique only works well for vertical posts~\cite{Buzea2005,Kaminska2005,Krause2010}. Other reported methods are, for example, x-ray reflectometry~\cite{Asgharizadeh2009} and gas adsorption isotherms~\cite{Krause2011}. Unfortunately, for samples with very large void fractions (deposition angles $>70^\circ$) the gas adsorption isotherm analysis fails because pores are not completely filled anymore due to bulk solidification~\cite{Krause2011}.

Generalized ellipsometry has been shown to be an excellent optical technique to determine anisotropic optical properties of complex nanostructured thin films and derive structural parameters such as thickness and void fraction from best-match model analysis~\cite{Kaminska2005,Beydaghyan2005,Nerbo2010}. Generalized ellipsometry is also capable of determining multiple film constituents within slanted columnar thin films. This has been recently shown for thin conformal passivation layers grown by atomic layer deposition and in-situ quantification of organic adsorbate attachment analysis~\cite{Schmidt2012a,Rodenhausen2012}.

However, since ellipsometry is an indirect measurement technique, adequate optical models have to be chosen to evaluate experimental data in order to obtain reliable optical and structural properties of anisotropic samples. The film structure of slanted columnar thin films, which are in general homogeneous anisotropic lossy composite materials consisting of slanted columns of regular shape and common orientation, induces form-birefringence and dichroism. Appropriate mixing formulas and effective medium homogenization approaches need to be applied to calculate an effective anisotropic dielectric medium response that renders the effects of the measured anisotropy~\cite{Shivolabook}.

In case of a biaxially anisotropic composite material, the classic ellipsometry model approach is to individually determine the three major axes dielectric functions without any implications on the kind of constituents and constituent fractions of the composite. This homogeneous biaxial layer approach can deliver structural information from a thickness parameter and Euler angles~\cite{Schmidt2009a,Schmidt2009b}. If constituent fractions and information about the shape of the constituents are desired results, a homogenization approach can be applied such that the three major axes dielectric functions can be constructed from a composite model that describes the effects of shape, average constituent fractions, and the use of constituent bulk-like optical constants for the materials of the buildings blocks (in general ellipsoidal inclusions).

The Bruggeman formalism is a homogenization approach with absolute equality between the constituents in a mixture, and was originally developed for a medium comprising two randomly distributed spherical dielectric components~\cite{Bruggeman1935}. This isotropic Bruggeman formula has been extensively discussed and generalized to treat materials with multiple anisotropic constituents by introducing so-called depolarization factors, which are functions of the shape of the inclusions~\cite{Polder1946,Stroud1975,Granqvist1977,Mackay2012}. For ellipsoidal particles however, two different modifications of the Bruggeman formalism were suggested, which differ in the definition of these depolarization factors.

The objective of this paper is a thorough comparison between both generalizations of the Bruggeman formalism and the homogeneous biaxial layer approach applied to slanted columnar thin films from different materials. Therefore, two different analysis procedures are presented to elucidate the potential of the two anisotropic Bruggeman models. All three model approaches have been employed to match experimentally acquired Mueller matrix spectra within the visible and near-infrared spectral regions from three different slanted columnar thin films made by glancing angle electron-beam deposition from cobalt, titanium, and silicon. Best-match model results are also compared to estimates from scanning electron microscopy images.


\section{Generalized Ellipsometry}
Generalized ellipsometry (GE), a non-destructive and non-invasive optical technique, has proven to be highly suitable for determining optical and structural properties of highly anisotropic nanostructured films from metals such as slanted columnar thin films or helical (chiral) sculptured thin films~\cite{Schmidt2009a,Schmidt2013ellinanoscale}. Measurements of the complex ratio $\rho$ of the $s$- and $p$-polarized reflection coefficients are presented here in terms of the Stokes descriptive system, where real-valued Mueller matrix elements $M_{ij}$ connect the Stokes parameters before and after sample interaction~\cite{HOE,Schubert2006a}.
The linear polarizability response of a nanostructured thin film due to an electric field $\mathbf{E}$ is a superposition of contributions along certain directions: ${\bf P}= \varrho_{a}(\bf{a}\cdot\bf{E})\bf{a} + \varrho_{b}(\bf{b}\cdot\bf{E})\bf{b} +\varrho_{c}(\bf{c}\cdot\bf{E})\bf{c}$~\cite{Dressel2008}.
In the laboratory Cartesian coordinate system the slanted columnar thin film is described by the second rank polarizability tensor $\boldsymbol{\chi}$ and $\mathbf{P} = (\boldsymbol{\varepsilon} - 1) \mathbf{E}=\boldsymbol{\chi} \mathbf{E}$. The Cartesian coordinate system $(x,y,z)$ is defined by the plane of incidence $(x,z)$ and the sample surface $(x,y)$. This Cartesian frame is rotated by the Euler angles ($\varphi, \theta, \psi$) to an auxiliary system $(\xi, \eta, \zeta)$ with $\zeta$ being parallel to $\bf{c}$~\cite{HOE,Schubertbook}. For orthorhombic, tetragonal, hexagonal, and trigonal systems a set of $\varphi, \theta, \psi$ exists with $\boldsymbol{\chi}$ being diagonal in $(\xi, \eta, \zeta)$. For monoclinic and triclinic systems an additional projection operation ${\bf U}$ onto the orthogonal auxiliary system $(\xi, \eta, \zeta)$ is necessary, which transforms the virtual orthogonal basis into a non-Cartesian system~\cite{Graefbook}:
\begin{equation}
\label{eqi5}\mathbf{U}=\left( \begin{array}{ccc}
\sin{\alpha} &  \frac{\cos{\gamma}-\cos\beta \cos\alpha}{\sin\alpha} & 0\\
0 & [\sin^2{\beta}- (\frac{\cos{\gamma} - \cos\beta\cos\alpha}{\sin{\alpha}})^2]^{\frac{1}{2}} & 0\\
\cos\alpha & \cos\beta & 1\\
\end{array} \right).
\end{equation}
Additional internal angles $\alpha, \beta, \gamma$ are introduced into the analysis procedure, and which differentiate between orthorhombic ($\alpha=\beta= \gamma=90^{\circ}$), monoclinic ($\beta \ne 90^{\circ}$), or triclinic ($\alpha \ne \beta \ne \gamma$) biaxial optical properties.

Ellipsometric data analysis for anisotropic thin film samples requires nonlinear regression methods, where measured and calculated GE data are matched as close as possible by varying appropriate physical model parameters~\cite{HOE,Schubertbook}. The quality of the match between model and experimental data can be measured by the mean square error (MSE)~\cite{Johs1993}.
The major axes polarization response functions $\varrho_a, \varrho_b, \varrho_c$ can be extracted on a wavelength-by-wavelength basis, i.e., without physical lineshape implementations and Kramers-Kronig consistency tests can then be done individually for dielectric functions along each axis~\cite{Dressel2008}. However, a generally more robust procedure is matching parameterized model dielectric functions to experimental data simultaneously for all spectral data points. Parametric models further prevent wavelength-by-wavelength measurement noise from becoming part of the extracted dielectric functions and greatly reduce the number of free parameters.

\section{Homogeneous Biaxial Layer Approach}
The homogeneous biaxial layer approach (HBLA) assumes that a given composite material can be described as a homogeneous medium whose anisotropic optical properties are rendered by a spatially constant dielectric function tensor. This dielectric function tensor must be symmetric since no magnetic or other non-reciprocal effects are considered. The dielectric function tensor, in general, comprises three frequency-dependent effective major axes dielectric functions $\varepsilon_j=1+\varrho(\omega)_{j}$ as described in Eq.~\ref{eq:monoclinicepstensor}, and may represent an anisotropic material resembling either orthorhombic, monoclinic, or triclinic optical symmetries.

Applied to a slanted columnar thin film, the optical equivalent can be, in the most simple case, a single biaxial layer described by the HBLA. This biaxial layer comprises then an optical thickness $d$, corresponding to the actual thickness of the nanostructured thin film as well as external Euler angles ($\varphi$, $\theta$, $\psi$) and internal angles ($\alpha$, $\beta$, $\gamma$) determining the orientation of the columns and sample during a particular measurement and biaxial properties, respectively (``structural parameters'')~\footnote{It is not a priori knowledge that the Euler angles, which diagonalize the HBLA tensor are equivalent to the intrinsic structural properties of the composite material such as slanting angle of the columns and orientation of the slanting plane relative to the external laboratory coordinate system. It has been confirmed by extensive investigations that Euler angles $\theta$ and $\varphi$ are identical to these properties.}. Furthermore, the three independent, complex, and wavelength-dependent functions $\varrho(\omega)_{j}$, pertinent to major polarizability axes $j=\mathbf{a},\mathbf{b},\mathbf{c}$ are referred to as ``optical properties''~\cite{Schmidt2009a,Schmidt2009b,Schmidt2013ellinanoscale}.

Explicitly, the dielectric tensor $\boldsymbol{{\varepsilon}}_{\textrm{t}}$ for a triclinic material takes the form
\begin{equation}\label{eq:monoclinicepstensor}
{\boldsymbol{\varepsilon}}_{\textrm{t}}=\mathbf{A}\mathbf{U} \left(
\begin{array}{ccc}
{\varrho(\omega)_{a} } & 0  & 0 \\
0  & {\varrho(\omega)_{b} } & 0 \\
0  & 0 & {\varrho(\omega)_{c} } \\
\end{array} \right) \mathbf{U}^{\rm t}\mathbf{A}^{\rm t},
\end{equation}
where $\mathbf{A}$ is the real-valued Euler angle rotation matrix and $\mathbf{U}$ is the projection matrix~\cite{Schmidt2013ellinanoscale}. Note that here the superscript ``t'' refers to the transpose of the respective matrix.

The HBLA does not allow to determine fractions of constituents within the composite material, nor the constituent bulk-like optical properties of the building blocks. However, the HBLA has several advantages over other effective medium approximations: (i) no initial assumptions such as optical properties of the constituents or material fractions are necessary, (ii) it is valid for absorbing and non-absorbing materials, and (iii) it does not depend on the structure size. Note that the actual structure size is disregarded in this homogenization approach. This procedure is considered valid since the dimensions and especially the diameter of the nanostructures under investigation are much smaller than the probing wavelength. Care must be taken when properties at shorter wavelengths are evaluated, because diffraction and scattering phenomena may be present.

In general, it is presumed that the HBLA method together with the assumption of one effective optical thickness $d$ applied to match experimental data for a slanted columnar thin films delivers the best possible dielectric tensor $\boldsymbol{\varepsilon}$, i.e. $\varepsilon_{\textrm{eff},j}$ are considered the true effective major axes dielectric functions and therefore target functions for other effective medium approximations.

%
%
%
%

\section{Bruggeman Formalisms}
The generalization of the Bruggeman formalism with a definition of the depolarization factors introduced to optics by Polder and van Santen (Eq.~\ref{eq:biaxBrugge1b}) has been extensively used and applied to the analysis of experimentally acquired data of anisotropic thin films~\cite{Polder1946,Granqvist1977,Smith1989,Mbise1997,Beydaghyan2005,Nerbo2010,Hofmann2011,Wakefield2011,Schmidt2012a,Schmidt2013ellinanoscale}. This formalism will be called henceforth ``traditional anisotropic Bruggeman effective medium approximation'' (TAB-EMA). The implicit TAB-EMA formulae for the three effective major dielectric functions $\varepsilon_{\stext{{eff},\textit{j}}}^{\stext{T}}$ with $j=a,b,c$ for a mixture of $m$ constituents with fractions $f_n$ and constituents bulk-like dielectric functions $\varepsilon_{{\rm c},n}$ are
\begin{equation}
\label{eq:biaxBrugge1b}
\sum\limits_{n = 1}^m f_n\frac{\varepsilon_{{\rm c},n}-\varepsilon_{\stext{{eff},\textit{j}}}^{\stext{T}}}{\varepsilon_{\stext{{eff},\textit{j}}}^{\stext{T}} + L^{\stext{D}}_{\stext{\textit{j}}}(\varepsilon_{{\rm c},n} - \varepsilon_{\stext{{eff},\textit{j}}}^{\stext{T}})}=0,
\end{equation}
with the depolarization factors
\begin{equation}
\label{eq:Brugge1depol}
L^{\stext{\rm D}}_{j}=\frac{U_xU_yU_z}{2}\int\limits_0^\infty\frac{(s+U_j^2)^{-1}{\rm d}s}{\sqrt{(s+U_x^2)(s+U_y^2)(s+U_z^2)}}.
\end{equation}
The definition of $L^{\stext{\rm D}}_{j}$ is based on the potential of uniformly polarized ellipsoids and has been adapted from magnetostatic theory where these parameters are well-known under the name demagnetizing factors~\cite{Osborn1945}. It is important to note that the real-valued depolarization factors $L^{\stext{\rm D}}_{j}$ only depend on the real-valued shape parameters $U_j$ of the ellipsoid and that the two ratios ($U_x/U_z$) and ($U_y/U_z$) serve to define the shape exactly. It can be shown that the depolarization factors of an ellipsoid satisfy the relation
\begin{equation}
\label{eq:unity}
L^{\stext{\rm D}}_{x}+L^{\stext{\rm D}}_{y}+L^{\stext{\rm D}}_{z}=1.
\end{equation}
Furthermore, the sum of all $f_n$ has to equal unity.
Analytical solutions for Eq.~\ref{eq:biaxBrugge1b} still exist even with several constituents $m$ and the physically correct solution of the polynomial equation can be determined by an algorithm based on conformal mapping, for example~\cite{Jansson1994}.

The second existing Bruggeman formalism comprises depolarization factors that are based on Green functions and was first introduced by Stroud in 1975~\cite{Stroud1975}. Recently, Mackay and Lakhtakia published explicit equations for these depolarization factors for the case of anisotropic inclusions~\cite{Mackay2012}. The effective permittivity parameters $\varepsilon_{\stext{{eff},\textit{j}}}^{\stext{R}}$ are given implicitly by the three coupled equations
\begin{equation}
\label{eq:biaxBrugge2b}
\sum\limits_{n = 1}^m f_n\frac{\varepsilon_{{\rm c},n}-\varepsilon_{\stext{{eff},\textit{j}}}^{\stext{R}}}{1 + D^{\stext{D}}_{\stext{\textit{j}}}(\varepsilon_{{\rm c},n} - \varepsilon_{\stext{{eff},\textit{j}}}^{\stext{R}})}=0,
\end{equation}
with the depolarization factors specified by the double integrals
\begin{subequations}\label{eq:Brugge2depola}
\begin{align}%
D^{\stext{\rm D}}_{x}&=\frac{1}{4\pi}\int\limits_{0}^{2\pi}\int\limits_{0}^{\pi}\frac{\sin^3\vartheta\cos^2\phi}{U_x^2\rho}{\rm d}\vartheta{\rm d}\phi\label{first},\\
D^{\stext{\rm D}}_{y}&=\frac{1}{4\pi}\int\limits_0^{2\pi}\int\limits_0^{\pi}\frac{\sin^3\vartheta\sin^2\phi}{U_y^2\rho}{\rm d}\vartheta{\rm d}\phi\label{second},\\
D^{\stext{\rm D}}_{z}&=\frac{1}{4\pi}\int\limits_0^{2\pi}\int\limits_0^{\pi}\frac{\sin\vartheta\cos^2\phi}{U_z^2\rho}{\rm d}\vartheta{\rm d}\phi\label{third},
\end{align}
\end{subequations}
which involve the scalar parameter
\begin{equation}
\label{eq:Brugge2rho}
\rho=\frac{\sin^2\vartheta\cos^2\phi}{U_x^2}\varepsilon_{\stext{{eff},\textit{x}}}^{\stext{R}}+\frac{\sin^2\vartheta\sin^2\phi}{U_y^2}\varepsilon_{\stext{{eff},\textit{y}}}^{\stext{R}}+\frac{\cos\vartheta}{U_z^2}\varepsilon_{\stext{{eff},\textit{z}}}^{\stext{R}}.
\end{equation}
The depolarization factors $D^{\stext{\rm D}}_{j}$ are, in general (lossy medium), complex parameters and are a function of the shape parameters $U_j$ of the ellipsoid as well as the effective permittivities $\varepsilon_{\stext{{eff},\textit{j}}}^{\stext{R}}$ of the medium. This formalism will be called henceforth ``rigorous anisotropic Bruggeman effective medium approximation'' (RAB-EMA). Note that due to the coupled nature of the RAB-EMA formalism generally numerical methods are necessary to calculate the effective permittivities $\varepsilon_{\stext{{eff},\textit{j}}}^{\stext{R}}$. In contrast to the TAB-EMA, the RAB-EMA has only been discussed mathematically and no reports on the application to evaluate experimentally acquired data from anisotropic samples exist. Furthermore, it should be noted that both theories (i) presume structural equivalence for all $m$ constituents and (ii) are identical for the limiting case of isotropic spherical inclusions ($U_x=U_y=U_z$).

\section{Analysis Procedures}
As shown previously, the optical response of slanted columnar thin films as the ones discussed here can be very well described using the HBLA method and assuming a single homogeneous biaxial layer~\cite{Schmidt2009a,Schmidt2009b,Schmidt2013ellinanoscale}. The latter assumption is also made for the TAB-EMA and RAB-EMA models. Furthermore, for both EMA models a composite material with two different constituents ($m=2$), one of them being void ($\varepsilon_{{\rm c},1}=1$), is assumed throughout the manuscript.
Two analysis procedures can be identified and results for both are presented and discussed below. The initial step, equal for both procedures, is the best-match model analysis with the HBLA method where experimentally acquired Mueller matrix spectra are matched as close as possible by varying structural and optical model parameters. First, wavelength-dependent functions $\varrho(\omega)_{j}$ pertinent to major polarizability axes $j=\mathbf{a}, \mathbf{b}, \mathbf{c}$ have been matched on a wavelength-by-wavelength basis. Subsequently, parameterized model dielectric functions have been implemented, also to warrant Kramers-Kronig consistency, and the best-match model calculation procedure was repeated for final results. Note that in accordance with previous findings the monoclinic angle $\beta$ was ranged for the model analysis such that $\beta\leq90^\circ$ and $\alpha=\gamma=90^\circ$. The monoclinic behavior is due to the specific thin film geometry and based on dielectric polarization charge coupling effects across neighboring slanted columns; i.e. a conducting wetting layer but electrically isolated nanocolumn tips~\cite{Schmidt2009a,Schmidt2009b,Schmidt2013ellinanoscale}. The subsequent two routes differ in what are considered target values for the best-match model regression analysis.

For analysis procedure I, the target values are the HBLA optical properties, that is, the effective major dielectric functions $\varepsilon_{\textrm{eff},j}$ and not the experimental data. Both TAB-EMA ($\varepsilon_{\stext{{eff},\textit{j}}}^{\stext{T}}$) and RAB-EMA ($\varepsilon_{\stext{{eff},\textit{j}}}^{\stext{R}}$) are matched to the target values by varying shape parameters $U_x$ and $U_y$ ($U_z=1$), void fraction parameter $f_1$ ($f_2=1-f_1$) as well as the wavelength-dependent constituent bulk-like dielectric function $\varepsilon_{{\rm c},2}$. The significance of this procedure is that the film thickness $d$ as well as the external Euler (rotations) and internal angles (projections), needed to diagonalize the HBLA tensor, do not affect the TAB- and RAB-EMA approaches. Note that any uncertainty of the HBLA results will also affect the subsequent EMA approaches. The quality of the regression analysis can be evaluated by comparing the weighted test function $\xi_{\varepsilon}$ ($\varepsilon$-MSE) values for both EMA approaches,
\begin{equation}\label{eq:MSE-EMA}
\xi_{\varepsilon} =\left[ \frac{1}{3S - K}\sum\limits_{j = 1}^S \sum\limits_{k = 0}^2 {{\left( {\frac{\varepsilon_{k,j}^{\stext{HBLA}} - \varepsilon_{k,j}^{\stext{mod}} }{\sigma_{j}^{\varepsilon_{k}} }}\right)^2 }} \right]^{1/2},
\end{equation}
where $S$ denotes the total number of wavelengths at which data are measured, $K$ is the number of model parameters, $\varepsilon_{k,j}^{\stext{HBLA}}$ and $\varepsilon_{k,j}^{\stext{mod}}$ are the HBLA target and model calculated effective major dielectric functions, respectively, and $\sigma_{j}^{\varepsilon_{k}}$ are the generated standard deviations ($\sigma_{j}^{\varepsilon_{k}}=0.01\varepsilon_{k,j}^{\stext{HBLA}}$) necessary for the biased weighing~\cite{Jellison1991}.

The target values for analysis procedure II are directly the experimentally acquired Mueller matrix spectra. Hence, the TAB- and RAB-EMA results are independent of the HBLA results, which means that additionally to shape parameters ($U_x, U_y$), void fraction $f_1$, and constituent bulk-like dielectric function $\varepsilon_{{\rm c},2}$ each EMA model will also deliver structural parameters, thickness as well as Euler and internal angles. The weighted test function $\xi_{\stext{M}}$ (M-MSE) may serve as a relative measure of quality to evaluate and compare the three model approaches
\begin{equation}\label{eq:MSE-MM}
 \xi_{\stext{M}} =\left[ \frac{1}{11S - K}\sum\limits_{j = 1}^S \sum\limits_{k = 1}^3 \sum\limits_{l = 1}^4 {{\left( {\frac{M_{kl,j}^{\stext{exp}} - M_{kl,j}^{\stext{mod}} }{\sigma_{j}^{\stext{exp}} }}\right)^2 } }\right]^{1/2}.
\end{equation}
Here, $M_{kl,j}^{\stext{exp}}$ and $M_{kl,j}^{\stext{mod}}$ are the experimentally determined and model calculated Mueller matrix elements, respectively, measured at photon energy $E=\hbar \omega_{j}$, and $\sigma_{j}^{\stext{exp}}$ are the standard deviations obtained during the measurement.

\begin{figure}[tbp]
\centerline{\includegraphics[width=.7\columnwidth, clip, trim=0 0 0 0]{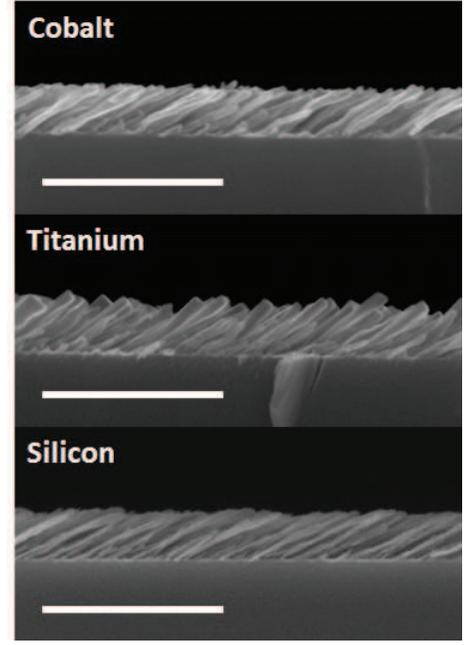}}
\caption{Cross-sectional high-resolution SEM images of the slanted columnar thin films from cobalt, titanium and silicon, respectively, on crystalline silicon substrates. Scale bars are 500~nm.}\label{fig:sem}
\end{figure}
\begin{figure}[tbp]
\centerline{\includegraphics[width=1.0\columnwidth, clip, trim=0 30 0 50]{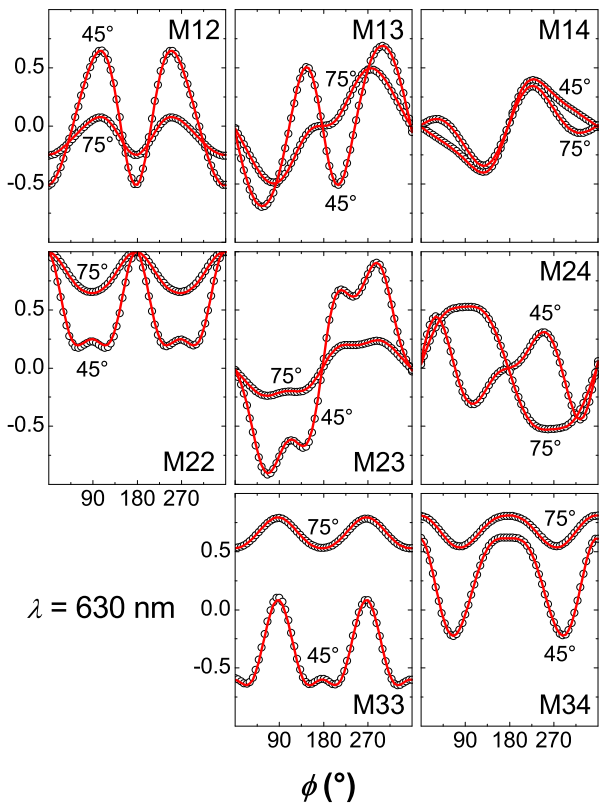}}
\caption{Experimental (circles) and best-match model calculated (solid lines) Mueller matrix data of a Si slanted columnar thin film versus sample azimuth angle $\phi$ at $\lambda = 630$~nm. The two graphs for each Mueller matrix element represent two different angles of incidence $\varPhi_{\rm a}=45^\circ$ and $\varPhi_{\rm a}=75^\circ$.}\label{fig:MMrot}
\end{figure}

\section{Experimental Details}
The slanted columnar thin films were deposited in a custom built ultra-high vacuum chamber by electron-beam evaporation of cobalt, titanium, and silicon pellets, respectively, onto Si(100) substrates with a native oxide layer of around 1.9~nm. The chamber background pressure was in the $10^{-9}$~mbar range and the substrates were held at room-temperature during the fabrication through a water-cooled sample holder. All three samples have been deposited at a constant particle flux of approximately 4~${\rm \AA}/$s measured at normal incidence while the substrate normal was tilted away from the particle flux by $85^\circ$. Figure~\ref{fig:sem} depicts cross-sectional high-resolution SEM images of the samples under investigation and Table~\ref{tab:sem-values} summarizes thickness and slanting angle determined by image analysis.

For each of the three materials, an approximately 100~nm thick solid reference film has been deposited at 2~${\rm \AA}/$s and at normal incident particle flux. To achieve a low surface roughness, the substrate has been rotated around its normal by 2~rpm during the growth.

\begin{table}[htbp]
  \centering
  \caption{Summary of thickness and slanting angle estimates from cross-sectional SEM image analysis for the three slanted columnar thin films depicted in Fig.~\ref{fig:sem}.}
    \begin{tabular}{lrrr}
    \hline\hline
    Parameter & Cobalt & Titanium & Silicon  \\
    \hline
    $t$ (nm)             & $115\pm5$   & $112\pm4$    & $108\pm3$ \\
    $\theta$ ($^\circ$)  & $65\pm3$    & $58\pm4$     & $62\pm2$  \\
    \hline\hline
    \end{tabular}%
  \label{tab:sem-values}%
\end{table}%

As soon as the samples had been taken out of the deposition chamber Mueller matrix ellipsometry spectra within a spectral range from 400 to 1650~nm have been acquired at angles of incidence $\varPhi_{\rm a}=45^\circ,55^\circ,65^\circ,75^\circ$ (M2000VI, J. A. Woollam Co. Inc.). Additionally, to allow for accurate evaluation of the anisotropy, at each angle $\varPhi_{\rm a}$ spectra were measured over a full azimuthal rotation every six degrees.

\section{Results}
\begin{figure}[tbp]
\centerline{\includegraphics[width=1.0\columnwidth, clip, trim=0 30 0 50]{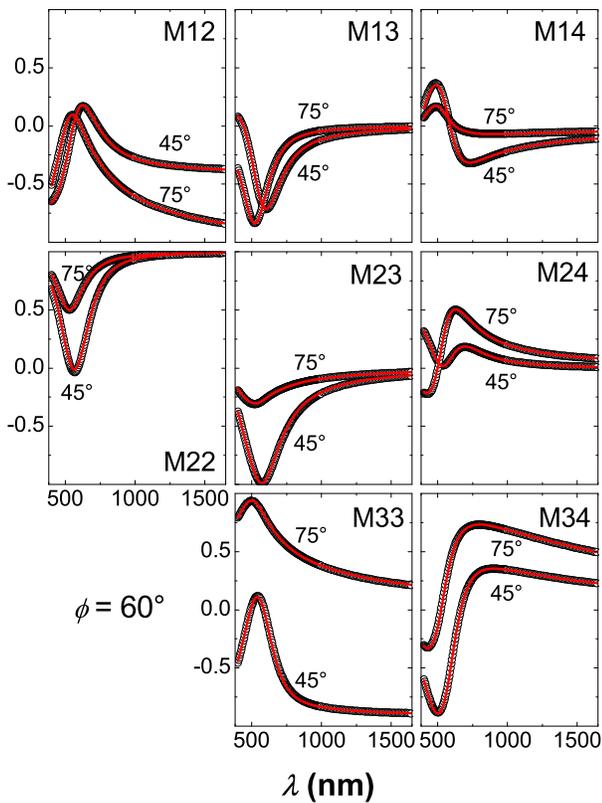}}
\caption{Experimental (circles) and best-match model calculated (solid lines) Mueller matrix data of a Si slanted columnar thin film for two different angles of incidence ($\varPhi_{\rm a}=45^\circ$ and $\varPhi_{\rm a}=75^\circ$) within the spectral range from 400 to 1650~nm.}\label{fig:MMwvl}
\end{figure}

\begin{figure*}[tbp]
\centerline{\includegraphics[width=.99\textwidth, clip, trim=0 0 0 0]{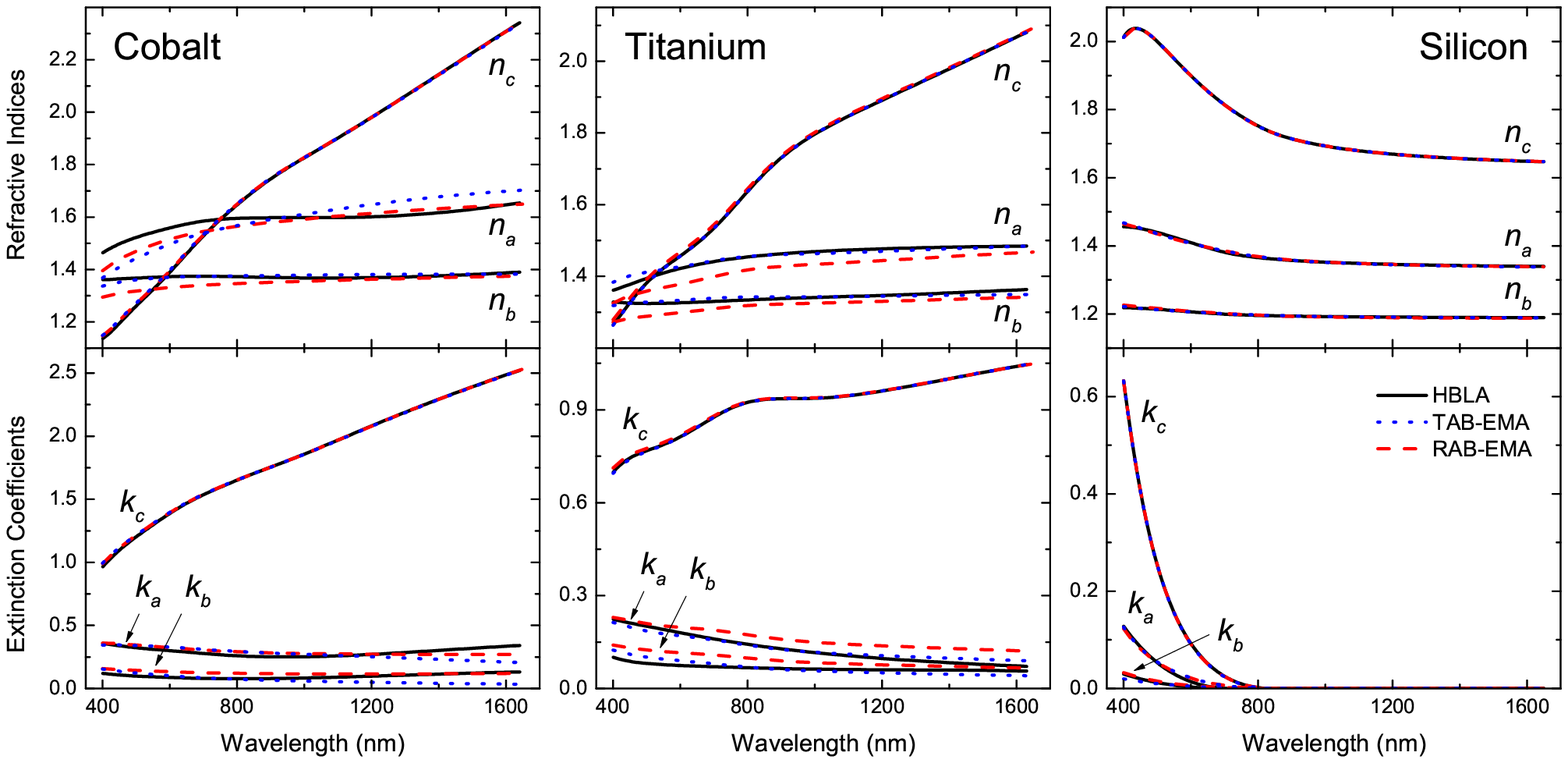}}
\caption{Effective major axes optical constants, refractive indices $n_j$ and extinction coefficients $k_j$, along major polarizability axes $\mathbf{a}$, $\mathbf{b}$, $\mathbf{c}$ of Co, Ti, and Si determined by HBLA (solid lines), TAB-EMA (dotted lines) and RAB-EMA (dashed lines). EMA results have been determined with analysis procedure I.}\label{fig:eff}
\end{figure*}

Figures~\ref{fig:MMrot} and~\ref{fig:MMwvl} depict azimuthal and spectral experimental and best-match model Mueller matrix data, respectively, for the Si sample and comparable graphs for the metal samples can be found in the literature~\cite{Schmidt2009a,Schmidt2009c,Schmidt2013ellinanoscale}. A summary of the best-match model structural parameters can be found in Table~\ref{tab:parameters} and effective major axes optical constants, $\varepsilon_{\stext{{eff},\textit{j}}}=(n_j+ik_j)^2$, are depicted in Fig.~\ref{fig:eff}. The Euler angle $\varphi$ depends merely on the azimuthal sample orientation $\phi$ during the ellipsometry measurement and has been omitted here; Euler angle $\psi$ was set to $0^\circ$ for all analysis procedures. Note that pseudo-isotropic orientations (sample azimuth positions with no mode conversion between \emph{p}- and \emph{s}-polarization states) can only exist if one major polarizability axes is within a plane, which is perpendicular to the plane of incidence ($\psi=0^\circ$)~\cite{Schmidt2013ellinanoscale,Schmidtdiss}.
For the two metal slanted columnar thin films, functions $\varrho(\omega)_{a}$, $\varrho(\omega)_{b}$, and $\varrho(\omega)_{c}$ have been parameterized with a combination of Lorentzian oscillators and for the Co thin film $\varrho(\omega)_{c}$ has an additional Drude term~\cite{HOE}. For the Si slanted columnar thin film each function $\varrho(\omega)_{j}$ comprises a single Tauc-Lorentz oscillator, which is the typical model dielectric function for amorphous Si~\cite{Jellison1996}.

Best-match model results determined with analysis procedure I are listed in Table~\ref{tab:method1} and the optical property results for all three slanted columnar thin films are shown in Fig.~\ref{fig:eff}. The corresponding wavelength-by-wavelength determined constituent bulk-like dielectric functions $\varepsilon_{{\rm c},2}=(n+ik)^2$ are depicted in Fig.~\ref{fig:bulk}.

Analysis procedure II results in best-match model structural parameters are listed in Table~\ref{tab:parameters}. The optical property results for all three slanted columnar thin films are shown in Fig.~\ref{fig:eff2} and the corresponding constituent bulk-like dielectric functions $\varepsilon_{{\rm c},2}=(n+ik)^2$ are depicted in Fig.~\ref{fig:bulk2}. Similarly to the HBLA, the bulk-like dielectric functions for the two metal slanted columnar thin films have been parameterized with a combination of Lorentzian oscillators and the model dielectric function for the Co slanted columns comprises an additional Drude term. For the Si slanted columns, $\varepsilon_{{\rm c},2}$ is composed of a single Tauc-Lorentz oscillator.

\section{Discussion}

\subsection{Optical Properties}
The effective major axes optical constants $n_j$ and $k_j$ of the Si slanted columnar thin film computed with TAB-EMA and RAB-EMA using analysis procedure I are in excellent agreement with the HBLA (Fig.~\ref{fig:eff}). While there are some deviations for the metal slanted columnar thin films regarding major polarizability axes $\mathbf{a}$ and $\mathbf{b}$ there is a very good match between all three model approaches along $\mathbf{c}$. This behavior is also reflected in the MSE values (Table~\ref{tab:method1}), which are lowest for the Si sample and considerably higher for the two metal samples. In particular, for the Ti nanostructured thin film the RAB-EMA formalism seems to underestimate $n_a$ and $n_b$ but overestimate $k_a$ and $k_b$. For the Co slanted columnar thin film, however, there are no obvious trends.

\begin{figure*}[tbp]
\centerline{\includegraphics[width=.99\textwidth, clip, trim=0 0 0 0]{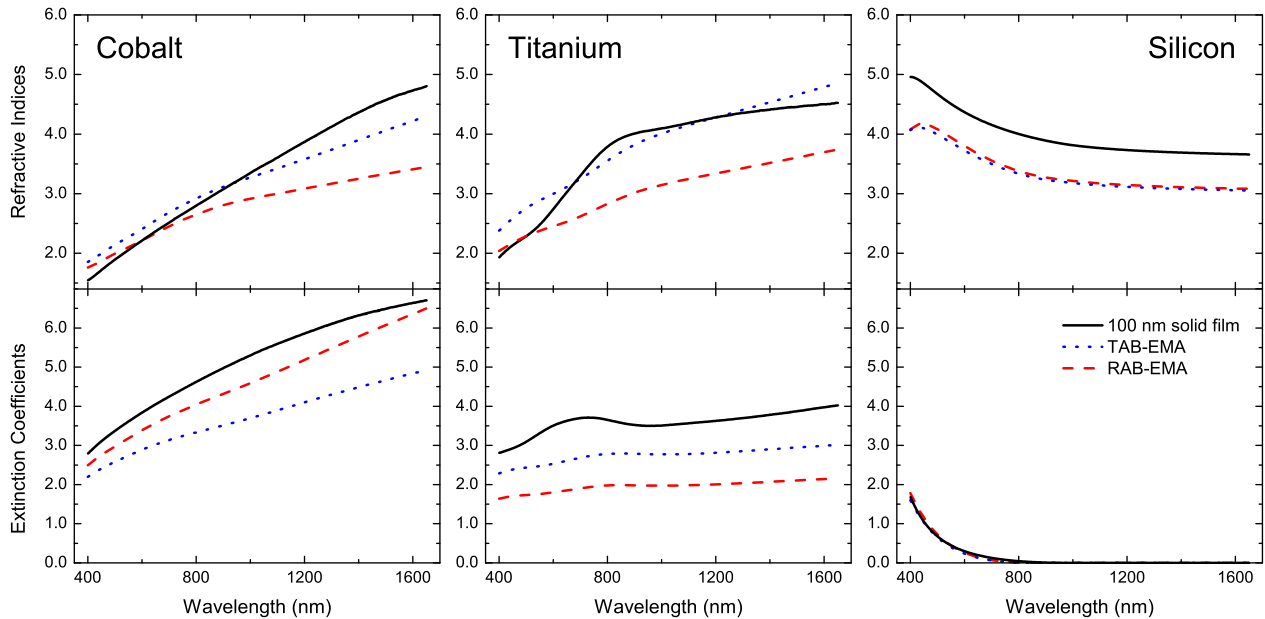}}
\caption{Constituent bulk-like optical constants, refractive indices and extinction coefficients, of Co, Ti, and Si as determined by TAB-EMA (dotted lines) and RAB-EMA (dashed lines), and also in comparison with reference data obtained from a 100~nm thick solid film (solid line). EMA results have been determined with analysis procedure I.}\label{fig:bulk}
\end{figure*}

The corresponding constituent bulk-like optical constants obtained by matching TAB-EMA and RAB-EMA to $\varepsilon_{\stext{{eff},\textit{j}}}^{\stext{HBLA}}$ are different from each other and also from the 100~nm thick solid film reference sample data (Fig.~\ref{fig:bulk}). In general, deviations from the bulk material optical constants are not unexpected since the optical properties of ultrathin films and nanoparticles may differ significantly due to surface and quantum confinement effects, for instance~\cite{Oates2004,Hövel2010,Alonso2010}.
No general trends can be observed between the two dissipative metal nanostructured samples within the measured spectral region, however, both TAB- and RAB-EMA agree very well for the Si slanted columnar thin film sample and only the refractive index is offset with respect to the reference sample data. This offset could stem from Si nanocrystals with different sizes present within the nanocolumns but not the purely amorphous solid reference sample~\cite{Alonso2010}. While the extinction coefficients for Ti exhibit substantial discrepancies the interband transition around 800~nm, even though slightly red-shifted with respect to the bulk reference data, is present in both Bruggeman approaches and has the same center energy ($1.52$~eV $\hateq$ 815~nm).

\begin{figure*}[tbp]
\centerline{\includegraphics[width=.99\textwidth, clip, trim=0 0 0 0]{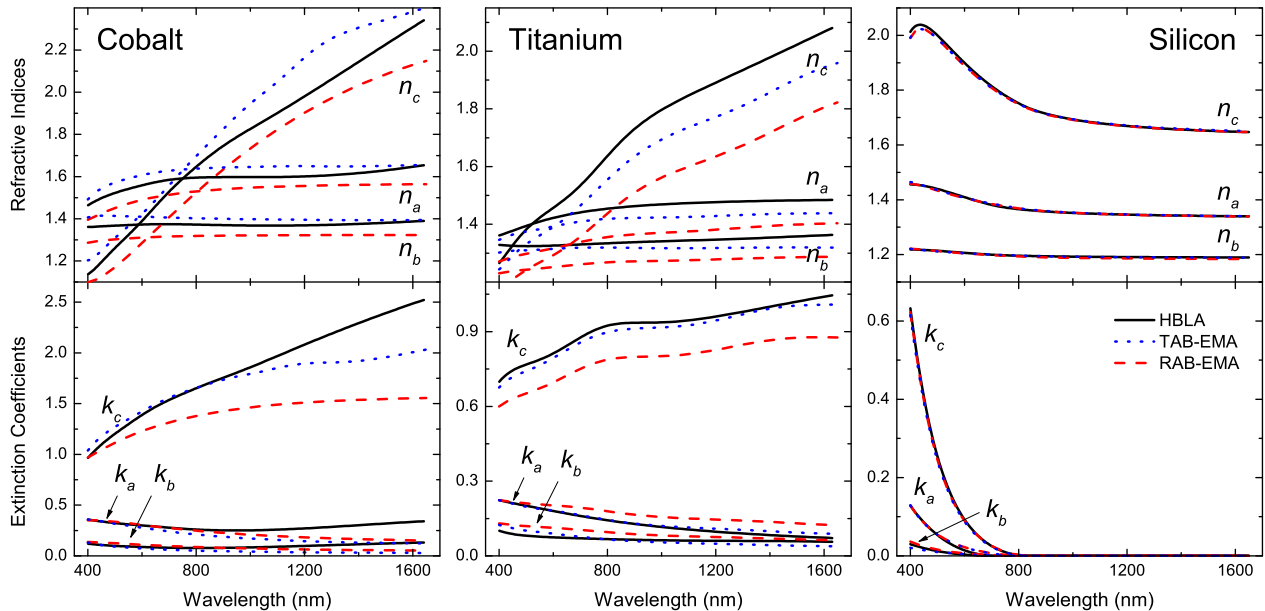}}
\caption{Effective major axes optical constants, refractive indices $n_j$ and extinction coefficients $k_j$, along major polarizability axes $\mathbf{a}$, $\mathbf{b}$, $\mathbf{c}$ of Co, Ti, and Si determined by HBLA (solid lines), TAB-EMA (dotted lines) and RAB-EMA (dashed lines). EMA results have been determined with analysis procedure II.}\label{fig:eff2}
\end{figure*}

Comparable to analysis procedure I, the TAB- and RAB-EMA effective major axes optical constants $n_j$ and $k_j$ of the Si slanted columnar thin film computed with analysis procedure II are in excellent agreement with the HBLA (Fig.~\ref{fig:eff2}). However, in contrast to analysis procedure I, the results obtained with the Bruggeman formalisms for both metal slanted columnar thin films deviate from the HBLA results along all three major polarizability axes. While for the Co sample, $n_j$ determined with the TAB-EMA and RAB-EMA overestimate and underestimate the corresponding HBLA indices, respectively, both EMA approaches underestimate the Ti HBLA $n_j$ with significant differences for the RAB-EMA. Exactly the same trends but with opposite directions can be observed for the film thicknesses. Overall, the TAB-EMA results, compared to the RAB-EMA, are slightly closer to $\varepsilon_{\stext{{eff},\textit{j}}}^{\stext{HBLA}}$ and also the MSE is reflecting this trend (Table~\ref{tab:parameters}). In general, it can be said that simply by considering the MSE the HBLA always delivers the best match between model and experimental data due to the independent determination of the effective optical constants along major polarizability axes. This observation is in accordance with the initial assumption that the HBLA method will deliver the best possible dielectric tensor.

\begin{figure*}[tbp]
\centerline{\includegraphics[width=.99\textwidth, clip, trim=0 0 0 0]{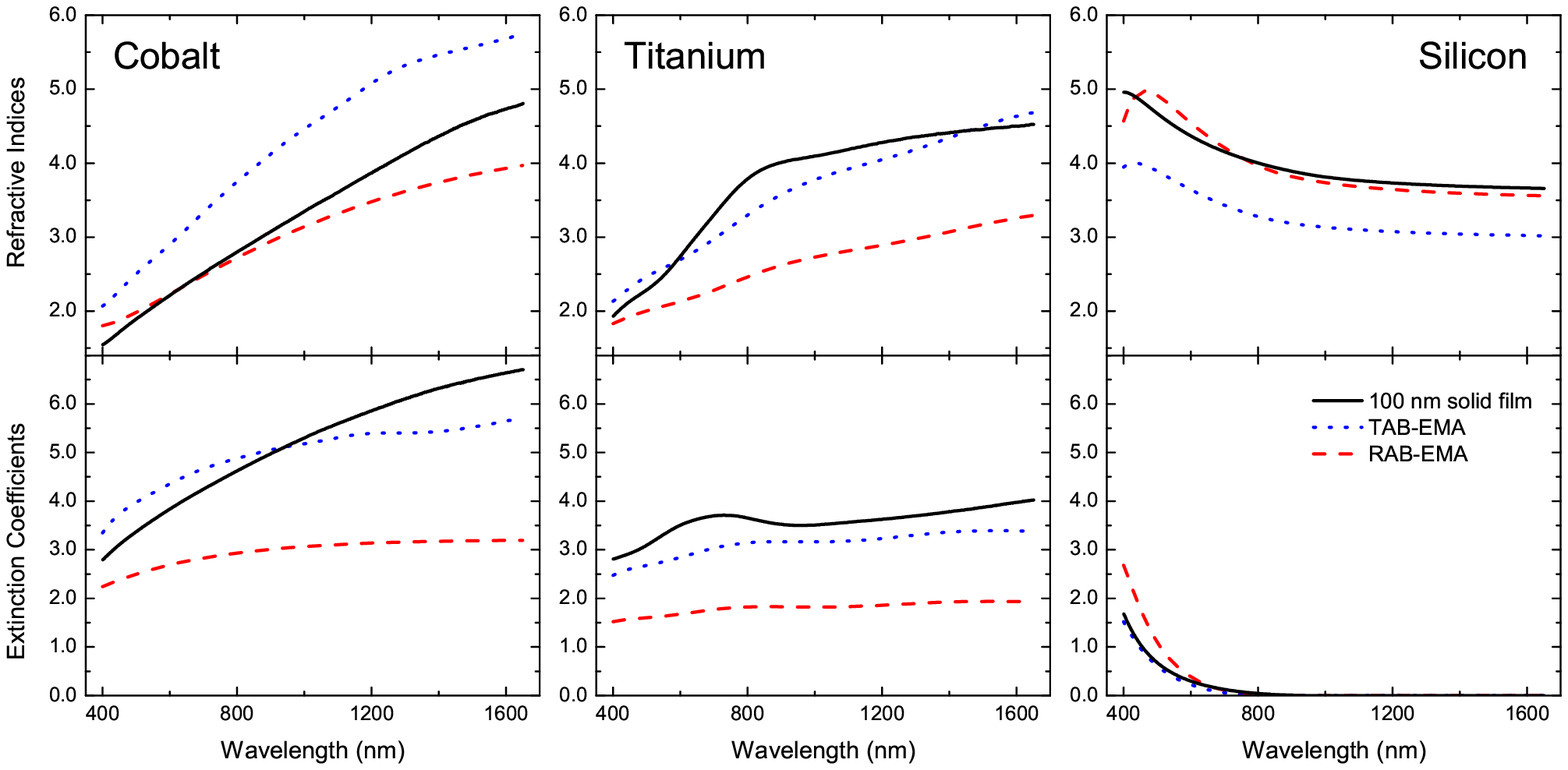}}
\caption{Constituent bulk-like optical constants, refractive indices and extinction coefficients, of Co, Ti, and Si as determined by TAB-EMA (dotted lines) and RAB-EMA (dashed lines), and also in comparison with data obtained from a 100~nm thick solid film (solid line). EMA results have been determined with analysis procedure II.}\label{fig:bulk2}
\end{figure*}

The corresponding constituent bulk-like optical constants resulting from matching TAB-EMA and RAB-EMA to experimental Mueller matrix spectra depicted in Fig.~\ref{fig:bulk2} are significantly different from each other, from data obtained from the 100~nm thin solid reference film, and also from the bulk-like optical constants determined with analysis procedure I (Fig.~\ref{fig:bulk}). Especially the RAB-EMA extinction coefficients for both Co and Ti nanostructured samples are substantially lower than the respective bulk reference data. Nevertheless, even though the extinction coefficients for Ti exhibit substantial discrepancies, the interband transition is present in both Bruggeman approaches and has the same center energy ($1.52$~eV $\hateq$ 815~nm), which is in agreement with analysis procedure I.

In summary, both analysis procedures reveal significant differences especially for the constituent bulk-like optical constants with respect to the solid reference sample as well as between analysis procedures, and also for the resulting major axes dielectric functions. The differences between bulk material reference data and constituent bulk-like optical constants of slanted columnar thin films determined with the Bruggeman EMAs discussed here may originate from a combination of circumstances. First of all, both Bruggeman formalisms are based on an idealized model of randomly distributed ellipsoidal particles and this description differs from the samples under investigation, which consist of columns with approximately elliptical cross-section attached to a substrate. Additionally, the optical model equivalent of a nanostructured thin film is a single anisotropic layer, which neglects non-idealities due to a ``surface roughness'' and a very thin nucleation layer. As mentioned above, another important consideration is that the dielectric properties of ultrathin metal films may differ from their respective bulk properties due to surface and quantum confinement effects, for example, which is very well possible here considering isolated columns with diameters of less than 20~nm~\cite{Hövel2010}. Besides that, since the acquisition of ellipsometric Mueller matrix spectra is performed in ambient conditions, a possible thin oxidation layer or even ambient humidity might alter the isotropic material optical constants slightly~\cite{Schmidt2012a}.

\subsection{Structural Parameters}
The film thicknesses $d$ of the absorbing metal slanted columnar thin films, optically determined with the HBLA and the TAB-EMA, are slightly lower compared to estimates obtained from cross-section SEM images while the RAB-EMA values agree fairly well (Tables~\ref{tab:sem-values} and~\ref{tab:parameters}). For the nanostructured thin film from Si however, all optical models are in excellent agreement and also with values obtained with the imaging technique.

All three model approaches are very reliable with respect to the determination of the columnar slanting angle $\theta$ (equivalent to the tilt of unit vector $\mathbf{c}$ from the surface normal towards the sample surface), which differs between models by $<2^\circ$ ($<1^\circ$ for the Si slanted columnar thin film) and is also in excellent agreement with SEM estimates.

Noticeable differences between the two EMA formalisms and the HBLA occur with respect to the monoclinic angle $\beta$. The Co slanted columnar thin film analysis reveals that the TAB-EMA result agrees well with the HBLA whereas the RAB-EMA suggests orthorhombic film properties. For the Ti sample however, all three model approaches result in monoclinic optical thin film properties with minor differences for the value of $\beta$. The semiconducting Si nanostructured film exhibits orthorhombic properties and only the TAB-EMA model results in a minor deviation from a right angle between axes $\mathbf{c}$ and $\mathbf{b}$. Note that $t$, $\theta$, and $\beta$ values are only computed for the two EMA model approaches with analysis procedure II.

The void fractions $f_1$ are results of each analysis procedure and both EMA formalisms deliver fairly consistent results (Tables~\ref{tab:method1} and~\ref{tab:parameters}). Analysis procedure I shows a maximum $f_1$ deviation between TAB and RAB of 2.5\% for the Co sample but the values for the Si sample are in excellent agreement. While the TAB-EMA void fractions for the metal slanted columnar thin films determined with analysis procedure II are slightly larger than the respective RAB-EMA values, the opposite is the case for the Si thin film. Trends are that the Si nanostructured sample exhibits a slightly larger void fraction compared to the metal samples deposited under identical conditions.
In general, void fraction values between 70\% and 80\% are also in agreement with general porosity trends for samples deposited at such glancing angles~\cite{Beydaghyan2005}. Unfortunately, no data for accurate direct comparison exist in the literature since either the method does not work for very large pore sizes (gas adsorption isotherm analysis)~\cite{Krause2011} or values have been determined by top-view SEM image analysis of vertically grown columnar thin films~\cite{Buzea2005,Kaminska2005}.

The most controversial results are the shape parameters ($U_x$, $U_y$), which should ideally render the geometry of the component particles of the homogenized medium. In the present case, the particles are rather columns composed of solid state matter and air, which will be approximated as ellipsoid by the Bruggeman formalisms. The fact that $U_z=1\neq U_x\neq U_y$ constitutes biaxial film properties and the larger $U_z$ is with respect to $U_x$ and $U_y$ the more elongated is the rendered ellipsoidal particle.

For analysis procedure I, only in the case of the Si slanted columnar thin film both EMA methods are consistent and close to the true column aspect ratio of about 10. For the Co sample the TAB-EMA and for the Ti sample the RAB-EMA substantially overestimate the aspect ratio, respectively. Interestingly, the $U_x/U_y$ ratios resulting from both Bruggeman approaches are fairly consistent for each slanted columnar thin film (Table~\ref{tab:method1}). Note that some shape parameters exhibit very large error bars due to parameter correlation, the ratio however can be determined with a very high precision by excluding one of them from the best-match model procedure.

Similar to analysis procedure I, the $U_x/U_y$ ratios, resulting from both Bruggeman approaches with analysis procedure II, are fairly consistent for each slanted columnar thin film. However, especially for the nanostructured samples from Ti and Co, ratios with respect to $U_z=1$ differ by one and two orders of magnitude, respectively. The RAB-EMA values are only close to the true column aspect ratio of about 10 in the case of the Si slanted columnar thin film. The TAB-EMA shape parameters are fairly consistent amongst the three samples under investigation and represent the actual column dimensions significantly better than the RAB-EMA values in case of the metal nanostructured thin films (Table~\ref{tab:parameters}).

In general, since the slanted columns interact with the substrate and with each other, an ellipsoidal shape is only a rough approximation. Hence $U_j$ should be considered as effective shape factors that are not necessarily representative for the true geometry of the inclusions~\cite{Granqvist1977}.

It should be noted that the RAB-EMA approach with its wavelength-dependent depolarization factors is not superior to the TAB-EMA formalism, which is possibly also due to the limited spectral range of investigation. However, the RAB-EMA might perhaps be the model of choice to analyze Mueller matrix data acquired over a very wide spectral range spanning from the visible all the way to the THz region, for example, since it has been shown recently that with the TAB-EMA depolarization factors and hence shape parameters may differ depending on the spectral range chosen for analysis~\cite{Hofmann2011}.

\begin{table}[tbp]
  \centering
  \caption{Summary of the best-match model parameters for Co, Ti, and Si slanted columnar thin films determined with analysis procedure I.}
    \begin{tabular}{ll|ccccc}
    \hline\hline
                           &     & Void \% &  $U_x$ & $U_y$ & $U_x/U_y$ & $\varepsilon$-MSE \\
    \hline
    \multirow{2}{*}{Co}   & TAB & 72.0(1)  &  0.07(1) & 0.05(1)  & 1.40& 0.1224  \\
                          & RAB & 74.5(1)  & 0.211(2) & 0.132(2) & 1.60& 0.0767  \\ \hline
    \multirow{2}{*}{Ti}   & TAB & 74.2(2)  & 0.208(5) & 0.180(5) & 1.16& 0.0236  \\
                          & RAB & 72.8(1)  & 0.005(9) & 0.003(7) & 1.40& 0.0736  \\ \hline
    \multirow{2}{*}{Si}   & TAB & 77.4(1)  & 0.223(6) & 0.118(4) & 1.89& 0.0052  \\
                          & RAB & 77.7(1)  & 0.189(7) & 0.088(4) & 2.15& 0.0053  \\
    \hline\hline
    \end{tabular}%
  \label{tab:method1}%
\end{table}%

\begin{table*}[tbp]
  \centering
  \caption{Summary of the best-match model parameters for Co, Ti, and Si slanted columnar thin films. EMA results have been determined with analysis procedure II.}
    \begin{tabular}{l|rrr|rrr|rrr}
    \hline\hline
              & \multicolumn{3}{c|}{Cobalt} &  \multicolumn{3}{c|}{Titanium} & \multicolumn{3}{c}{Silicon} \\
    Parameter & HBLA & TAB-EMA & RAB-EMA  & HBLA & TAB-EMA & RAB-EMA & HBLA & TAB-EMA & RAB-EMA  \\
    \hline
    $t$ (nm)            & 106.9(1)  & 101.51(7)   & 114.81(8)   & 100.8(1) & 103.0(2)    & 114.9(1)    & 108.8(3)  & 109.07(3)    & 109.31(3)    \\
    $\theta$ ($^\circ$) & 63.68(2)  & 62.92(2)    & 64.80(2)    & 57.63(4) & 56.95(3)    & 58.42(3)    & 60.94(4)  & 61.62(4)     & 61.10(3)     \\
    $\beta$ ($^\circ$)  & 83.7(1)   & 80.91(5)    & 90.0(1)     & 79.9(1)  & 81.23(6)    & 85.12(7)    & 90.0(1)   & 88.0(1)      & 90.0(1)      \\
    Void \%             & ---       & 75.37(1)    & 73.66(2)    & ---      & 78.26(4)    & 74.75(5)    & ---       & 77.13(4)     & 80.2(1)      \\
    $U_x$               & ---       & 0.406(1)    & 0.004(2)    & ---      & 0.380(3)    & 0.02(1)     & ---       & 0.198(3)     & 0.332(6)     \\
    $U_y$               & ---       & 0.311(1)    & 0.003(1)    & ---      & 0.312(3)    & 0.012(6)    & ---       & 0.103(3)     & 0.163(4)     \\
    $U_x/U_y$           & ---       & 1.31        & 1.72        & ---      & 1.22        & 1.46        & ---       & 1.98         & 2.04         \\
    M-MSE               & 7.7       & 12.5        & 19.73       & 6.2      & 10.23       & 12.8        & 8.02      & 10.46        & 10.51        \\
    \hline\hline
    \end{tabular}%
  \label{tab:parameters}%
\end{table*}%

\subsection{Applications}
The findings discussed in the above sections are relevant for a number of optical device structure applications based on a birefringence change principle. For example, detecting changes in the environmental conditions by monitoring the adsorption of molecules out of the gaseous phase such as optical humidity sensing or monitoring adsorption of organic molecules out of a liquid phase by means of generalized ellipsometry~\cite{Rodenhausen2012}. Furthermore, these model approaches are important to characterize changes in constituent fractions or other structural variations. Constituent fraction changes, for instance, occur when stimuli responsive functional polymer brushes coated onto slanted columnar thin films swell or deswell upon environmental changes or during conformal growth by atomic layer deposition~\cite{Kasputis2013,Schmidt2012a,Schmidt2013a}. Slanting angle and associated thickness and constituent fraction variations may appear upon infiltration of polymers by spin coating, for instance~\cite{Kasputis2013,Liang2013}.

\section{Conclusions}
Two different generalized Bruggeman formalisms to determine the effective macroscopic permittivity of a homogenized composite material have been applied to model ellipsometric Mueller matrix spectra. The obtained results have been discussed and compared to a homogeneous biaxial layer approach as well as estimates from cross-sectional electron microscopy images.
Since both anisotropic Bruggeman formalisms are approximations and have been developed for isolated, randomly distributed component particles of ellipsoidal shape, a slanted columnar thin film of any material is not a perfect match for this idealized model scenario. Especially, since the columns interact with the substrate and with each other, an ellipsoidal geometry is only a rough approximation.

Despite all the the above mentioned model non-idealities both EMA formalisms result in very good estimates for structural parameters such as thickness, slanting angle, and constituent fractions as analysis procedure II (match model to experimental Mueller matrix spectra) has revealed. However, shape parameters need to be considered as effective values, which do not necessarily represent the true structure geometry. For slanted columnar thin films of the kind presented here and investigated within a narrow spectral range, no preference can be given to any one of the Bruggeman EMA model approaches. For the metal slanted columnar thin films the TAB-EMA produces slightly smaller MSE values whereas both Bruggeman formalisms are equally good when matching data from a Si slanted columnar thin film.

However, if an accurate determination of the effective major dielectric functions is the desired result, the HBLA model needs to be used. Then by applying analysis procedure I (match model to HBLA effective major dielectric functions instead of experimental data), often desired constituent fractions can be obtained as well as the constituent bulk-like dielectric functions, which may be significantly different from the respective bulk material dielectric functions. For lossless or near lossless materials (extinction coefficient $k$ small or zero) both TAB-EMA and RAB-EMA deliver almost identical constituent bulk-like dielectric functions as shown in the case of Si slanted columnar thin films. For the lossy metal nanostructured samples however, both EMA approaches differ substantially and here probably the RAB-EMA should be given preference due to the rigorous electrodynamic approach.

\section{Acknowledgements}
D.S. would like to thank C.M. Herzinger (J.A. Woollam Co., Inc.) and T. Hofmann (University of Nebraska-Lincoln) for fruitful discussions. The authors acknowledge financial support from the National Science Foundation in RII (EPS-1004094), CAREER (ECCS-0846329), and MRSEC (DMR-0820521), the University of Nebraska-Lincoln, and the J.A. Woollam Foundation.


%

\end{document}